# Functional thin films as cathode/electrolyte interlayers: a strategy to enhance the performance and durability of solid oxide fuel cells

Marina Machado,[a] Federico Baiutti,[b] Lucile Bernadet,[b] Alex Morata,[b] Marc Nuñez,[b] Jan Pieter Ouweltjes,[c] Fabio Coral Fonseca,[a] Marc Torrell,[b] and Albert Tarancón[b,d*]

[a]IPEN-CNEN, Nuclear and Energy Research Institute, 05508-000, São Paulo, SP, Brazil.
[b]Catalonia Institute for Energy Research (IREC), Department of Advanced Materials for Energy, Jardins de les Dones de Negre, 1, 08930 Sant Adrià de Besòs, Barcelona, Spain.
[c]SOLIDPower SpA, Viale Trento 117, Mezzolombardo, 38017, Italy.
[d]ICREA, Passeig Lluís Companys 23, 08010 Barcelona, Spain.
[*]E-mail: atarancon@irec.cat

**Abstract**

Electrochemical devices such as solid oxide fuel cells (SOFC) may greatly benefit from the implementation of nanoengineered thin-film multifunctional layers providing, alongside enhanced electrochemical activity, improved mechanical, and long-term stability. In this study, an ultrathin (400 nm) bilayer of samarium-doped ceria and a self-assembled nanocomposite made of $Sm_{0.2}Ce_{0.8}O_{1.9}$-$La_{0.8}Sr_{0.2}MnO_{3-\delta}$ was fabricated by pulsed laser deposition and is employed as a functional oxygen electrode in an anode-supported solid oxide fuel cell. Introducing the functional bilayer in the cell architecture results in a simple processing technique for the fabrication of high-performance fuel cells (power density 1.0 W·cm$^{-2}$ at 0.7 V and 750 °C). Durability tests were carried out for up to 1500 h, showing a small degradation under extreme operating conditions of 1 A·cm$^{-2}$, while a stable behaviour at 0.5 A·cm$^{-2}$ (2.8% $V_{in}$·kh$^{-1}$). Post-test analyses, including scanning and transmission electron microscopy and electrochemical impedance spectroscopy, demonstrate that the nanoengineered thin film layers remain mostly morphologically stable after the operation.

**Introduction**

Solid oxide fuel cells (SOFCs) are highly efficient electrochemical devices that convert the chemical energy of a fuel directly into electricity.[1,2] These ceramic multilayer devices contain several interfaces, some of which are critical for the final performance of the cell. In particular, the oxygen electrode/electrolyte interface is responsible for the decrease of the cell's long-term performance, when the formation of secondary phases occurs as a consequence of cationic interdiffusion between commonly used yttria-stabilized zirconia (YSZ) electrolytes and state-of-the-art cobaltite-ferrite cathodes $La_{1-x}Sr_xCo_{1-y}Fe_yO_{3-\delta}$ (LSCF). The mixed ionic electronic conducting (MIEC) LSCF perovskites reacts with YSZ to form insulating phases such as $SrZrO_3$ and $La_2Zr_2O_7$.[3–5] Such undesired phases hinder interfacial fast mass and charge transfer necessary for high-performance SOFC at reduced operating temperature (<800 °C).[6,7] Ceria-based diffusion barrier layer (BL) is a widely employed engineering strategy to avoid the interdiffusion processes at the electrolyte/cathode interface.[8–13] Among the various fabrication techniques, the physical deposition of thin-film,[14,15] and in particular pulsed laser deposition (PLD), have shown to be highly promising for obtaining dense and continuous BLs with reduced thickness (typically between 500 nm and 2000 nm).[16–19] Morales et al.[20] have shown that a 2 µm-thick barrier leads to an improved performance (70% higher peak power density) compared to state-of-art cells, showing outstanding stability even under 5,000 h of continuous operation. In order to enhance the capability of PLD BLs to block cation diffusion, high-temperature post-annealing is typically employed for blocking diffusion pathways through grain boundaries.[21]

Besides the diffusion BL, the PLD technique has also been used to deposit other components, such as the electrodes and electrolyte as thin films of an SOFC.[22,23] Those strategies have shown to improve the performance of the cell and enable the production of small scale SOFC for portable devices. Recently, novel approaches of thin films based on nano-scaled structures have been applied between cathode and electrolyte to decrease the oxygen electrode resistance.[24–30] Su et al.[24] and Yoon et al.[25] have successfully grown nanocomposite structures of LSM/YSZ and LSCO/GDC, respectively, reporting an improvement in electrochemical properties. Develos-Bagarinao et al.[26] proposed a dense multilayer heteroepitaxial cathode of LSC and GDC with interface-induced enhancement of oxygen surface exchange and oxygen ion diffusivity resulting in a superior fuel cell performance. While fuel cells comprising nanocomposite layers may exhibit enhanced power density by improving adhesion properties, increasing the density of the active surface area to facilitate the migration of oxygen ions, charge transfer processes, and long-term thermochemical stability remain a challenge.[27,28,31] In this regard, a previous report investigated the fundamental electrochemical properties of LSM-SDC vertically aligned nanocomposites (VANs) deposited on model single crystal electrolytes.[32] These VANs showed a greatly improved long-term thermal stability with respect to state-of-the-art electrode materials by suppressing the Sr segregation to the surface, along with enhanced oxygen kinetics as a consequence of fast diffusion and incorporation pathways promoted by the microstructural alignment.[32] Moreover, unlike standard multilayer films, VANs are self-assembled structures, with stable interfaces formed at growth temperature, that exhibit good microstructural stability at operation temperatures.

In the present study, a multifunctional bilayer system composed of samarium doped-ceria BL (SDC, 200 nm thick) and a nanostructured composite functional layer (NFL) of $La_{0.8}Sr_{0.2}MnO_{3-\delta}$-$Sm_{0.2}Ce_{0.8}O_{1.9}$ VAN (LSM-SDC, 200 nm thick) is fabricated by a one-step PLD deposition in the cathode/electrolyte interface of solid oxide fuel cells. After microstructural optimization using symmetric cells, the BL and NFL are employed as a functional bilayer in state-of-the-art anode supported cells with LSCF cathode. A NFL containing LSM rather than LSCF was chosen to further inhibit the formation of resistive phases. Detailed electrochemical characterization of fuel cell, including long-term degradation measurements under high current, is carried out, revealing a strongly increased power density with good long-term stability. Advanced structural characterization and post-test analysis highlight the stability of the BL+NFL system, revealing an efficient role of this bilayer as a diffusion blocking layer after long-term operation.

**Experimental**

Symmetrical cells were prepared by deposition of thin layers by PLD on a 150 μm 8YSZ tape (Kerafol). Particularly, a bilayer formed by an SDC BL and a NFL of LSM-SDC was deposited by PLD in a multitarget chamber from PVD Products (PVD5000) with a KrF excimer laser ($\lambda$ = 248 nm). The targets used for this purpose were $Sm_{0.2}Ce_{0.8}O_{1.9}$ (Kceracell) for the pure SDC BL and 50:50 wt% $(La_{0.8}Sr_{0.2})_{0.98}MnO_3$ (Kceracell): $Sm_{0.2}Ce_{0.8}O_{1.9}$ (Kceracell) for the NFL target. An energy fluency of around 1 J·cm$^{-2}$ was used for the ablation of the targets and a frequency of 10 Hz for the BL and 2 Hz for the NFL. During the deposition process, the distance between the substrate and the target was 90 mm, the temperature of the chamber was kept at 750 °C under an oxygen partial pressure of 0.7 Pa. A cathode layer of $(La_{0.6}Sr_{0.4})_{0.97}Co_{0.2}Fe_{0.8}O_3$ (LSCF) (Kceracell) was deposited on top of the as-deposited PLD layers by a 3-axis automated airbrush (Print3D Solutions), using an ethanol-based dispersion ink. The evolution of the NFL layer and its interface with the other cell components was studied by direct deposition of the NFL on YSZ tape. For this purpose, a set of cells with only the NFL and another set with NFL and LSCF cathode were deposited and thermally treated at 900, 1000 and 1100 °C, to determine the optimal adhesion temperature. According to the EIS analysis of the symmetric cells, 1000 °C was defined as the optimal temperature (details supplied in the supplementary information section). Likewise, the samples with the bilayer, using BL+NFL as interlayers, were sintered at 1000 °C. The symmetric cells with the different interlayers and sintering temperatures were electrochemically characterized at temperature ranging from 550 to 800 °C under a synthetic air atmosphere.

Deposited samples (as-deposited and after heat treatment) were characterized by X-ray diffraction (XRD) on a Bruker-D8 Advance instrument using Cu-Kα radiation with a nickel filter and a Lynx Eye detector. The microstructural characterization of the cells before and after operation was obtained from lamellae of the samples prepared by focused ion beam scanning electron microscope (FIB-SEM) Dual Beam Helios 650. The SEM images were obtained by the Dual Beam Helios 650 SEM and transmission electron microscopy (TEM) analyses of the lamellae and EDX mapping were carried out using the JEOL 2010F TEM equipment.

Complete SOFC cells were prepared to evaluate the performance and the durability of a Ni-YSZ/YSZ/SDCBL/LSM-SDCNFL/LSCF single cell. A state-of-the-art anode supported half-cell supplied was used as substrate. The SDC barrier layer was deposited on top of the 8YSZ tape followed by the deposition of the LSM-SDC ceramic nanocomposite. Finally, a LSCF cathode, thickness of 18 µm, was airbrushed on an area of 1.54 cm$^2$ on top of the NFL. A ProboStat™ (NorECS) system placed inside a high-temperature tube furnace was used to test the developed cells under symmetric and under fuel cell conditions. Ceramabond™ (Aremco) paste was used to seal the cell on the test rig and separate anodic and cathodic chambers. The measurements of the complete cells were done under dry $H_2$ fuel in the anodic compartment and synthetic air and oxygen in the cathode with a flow density of 22.22 Nml·min$^{-1}$·cm$^{-2}$. The electrochemical characterization of the fuel cell was performed by impedance spectroscopy (EIS) and I-V measurements were carried out using the PARSTAT® 2273 potentiostat and the long-term durability test was done using an M9700 electronic load from Maynuo Electronic Co. Ltd. The EIS and I-V curves were obtained at OCV and with a current density of 0.5 A·cm$^{-2}$ at 750 °C. The long-term durability test was carried out for 1500 h at 750 °C with a current density of 1 A·cm$^{-2}$ in the first 250 h and a current density of 0.5 A·cm$^{-2}$ for the remaining 1250 h.

## Results and discussion

X-ray diffraction was used to confirm the crystalline phases of the thin functional layers of SDC (BL), LSM-SDC (NFL) and the bilayer SDC+LSM-SDC (BL+NFL) deposited by PLD on YSZ substrates (**Fig. 1**). The diffraction patterns revealed the cubic structure of both the YSZ substrate and of the SDC. The LSM diffraction peaks are clearly visible and match with pseudo-cubic structure typical of thin film LSM layers.[33] Samples treated at 1000 °C for the electrode adhesion (pl. see the **Supplementary Fig. S1**) show no significant changes on the XRD data, without formation of additional phases or significant peak shifts.

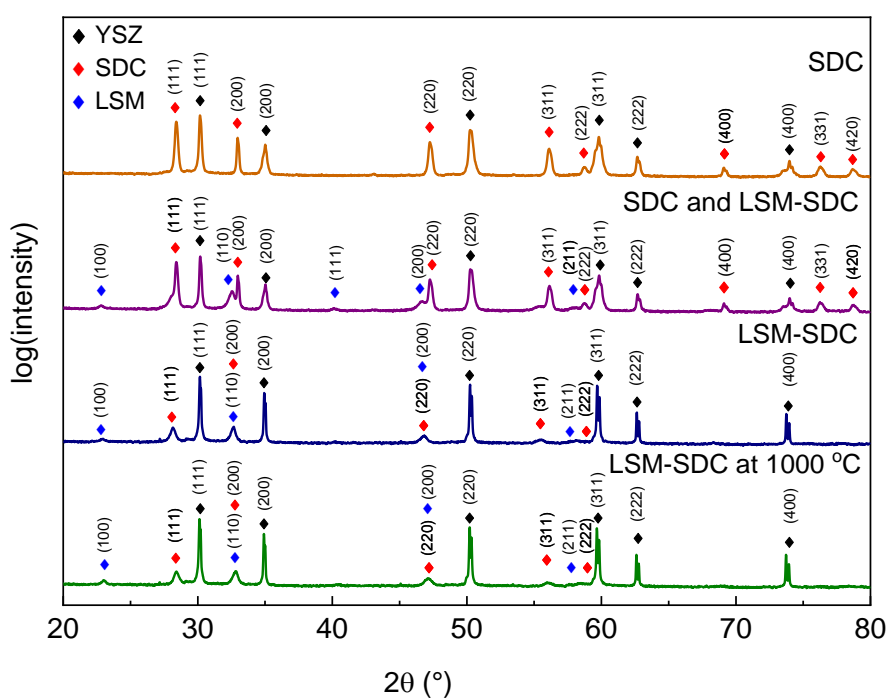

**Fig. 1** X-ray diffraction patterns of PLD layers on YSZ. Top to bottom: SDC, LSM-SDC and SDC+LSM-SDC as-deposited.

To determine the morphological characteristics of the electrolyte-electrode interface microstructure, SEM cross-section images were obtained as shown in **Fig. 2 (a)**. One can observe the BL+NFL sandwiched between the YSZ substrate and airbrushed LSCF layer. Please note that the original mesostructure was sintered at 1000 °C. The images show a 400 nm thick homogeneous bilayer of both SDC (200 nm) and LSM-SDC (200 nm). The PLD layers exhibit high density and are well-attached both to the YSZ electrolyte and to the LSFC cathode. The chemical composition was assessed by high-resolution EDX mapping, which was carried out on the cross-section area near the interface, as shown in **Fig. 2 (b)**. From the elemental analyses, the BL and the NFL layer are visibly distinguishable. It is also possible to identify a clear vertical periodicity, distinguishable by the signals of Ce and Sm (~ 200 nm column spacing) in the NFL layer, confirming the alternate vertical growth of SDC and LSM. Notably, unlike typical PLD nanocomposite model systems,[34] vertical alternation was here obtained without the need of a well-defined substrate orientation, as the BL is highly polycrystalline (cf. **Fig. 1**).

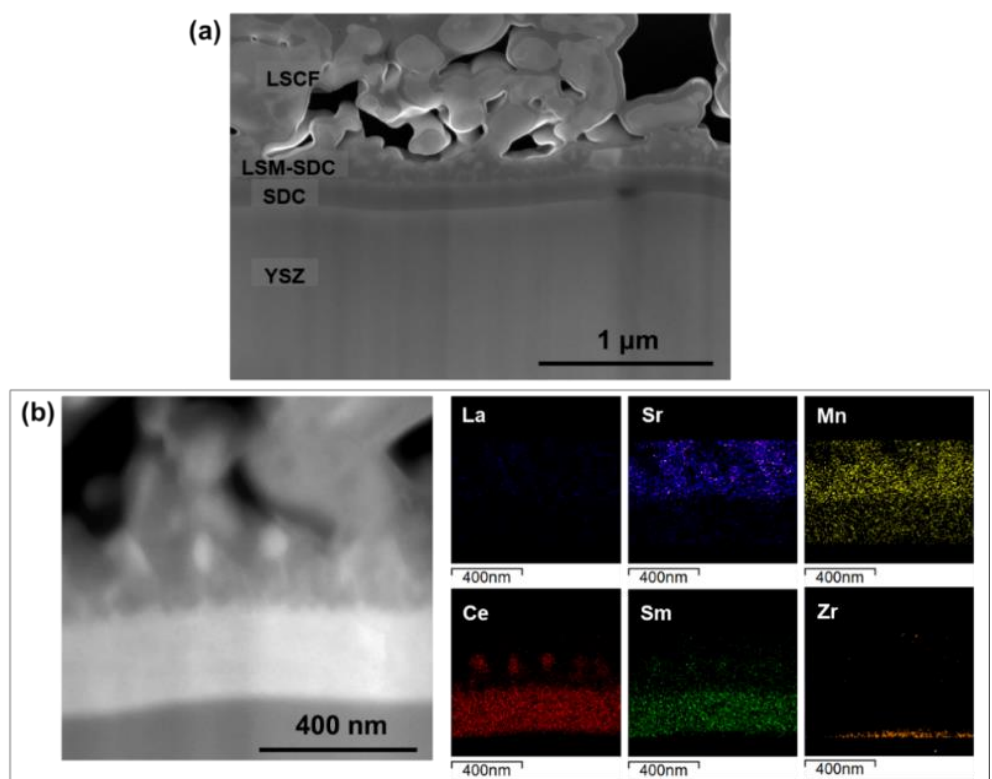

**Fig. 2 (a)** Cross-sectional SEM image of the SDC BL and the LSM-SDC NFL deposited on YSZ support with LSCF treated at 1000 °C. **(b)** STEM image of the cross-section with EDX mapping of the deposited thin films.

The electrochemical impedance spectroscopy (EIS) analyses were conducted on symmetrical cells with configuration LSCF/NFL/BL/YSZ/BL/NFL/LSCF. The EIS analysis of the symmetrical cell sintered at 1000 °C is shown in **Fig. 3**, such temperature was determined to be the optimal sintering temperature (Supplementary **Fig. S2 (a)**). The EIS data of the symmetrical cell was fitted with the equivalent circuit ($LR_S(R_{P1}CPE_{P1})(R_{P2}CPE_{P2})$), which is the simplest circuit able to fit the impedance contributions observed in the Nyquist plots.

Notably, the addition of the bilayer (BL+NFL shown in **Fig. 3 (a)**) was found to decrease the total resistance of the cell as compared with the sample with only the NFL (**Fig. S2 (a)**). The total electrical resistance at 750 °C decreased from 0.94 $\Omega \cdot cm^2$ with the NFL to 0.48 $\Omega \cdot cm^2$ with the bilayer BL+NFL (**Fig. S2 (b)**). The most important difference was noted on the polarization arc that appears at high frequency (RP1), which decreased from 0.42 $\Omega \cdot cm^2$ in the sample with the NFL to 0.061 $\Omega \cdot cm^2$ with the BL+NFL at 750 °C. Such high-frequency arc is usually attributed to interfacial impedance contributions at the electrode-electrolyte interface.[35,36] According to this interpretation, the presence of the BL favours the out-of-plane oxygen transport from the electrode to the electrolyte. Nyquist plot and area specific resistances (ASR) graph for the symmetrical cell without the interlayers is reported in Supplementary **Fig. S3**, exhibiting a more than 40-fold increase in ASR as compared to the cell with the bilayer.

**Fig. 3 (b)** shows the temperature dependence of the ASR components (series, polarization, and total) of the BL+NFL cell. The polarization resistances are comparable to experimental results reported for conventional LSCF cathodes.[29,37] A low activation energy is observed for both serial and polarization resistances. The calculated polarization activation energy is 1.23 eV (in comparison with ~ 1.4 eV for LSCF cathodes),[37–40] suggesting that the nanofunctional layer facilitates the ORR.

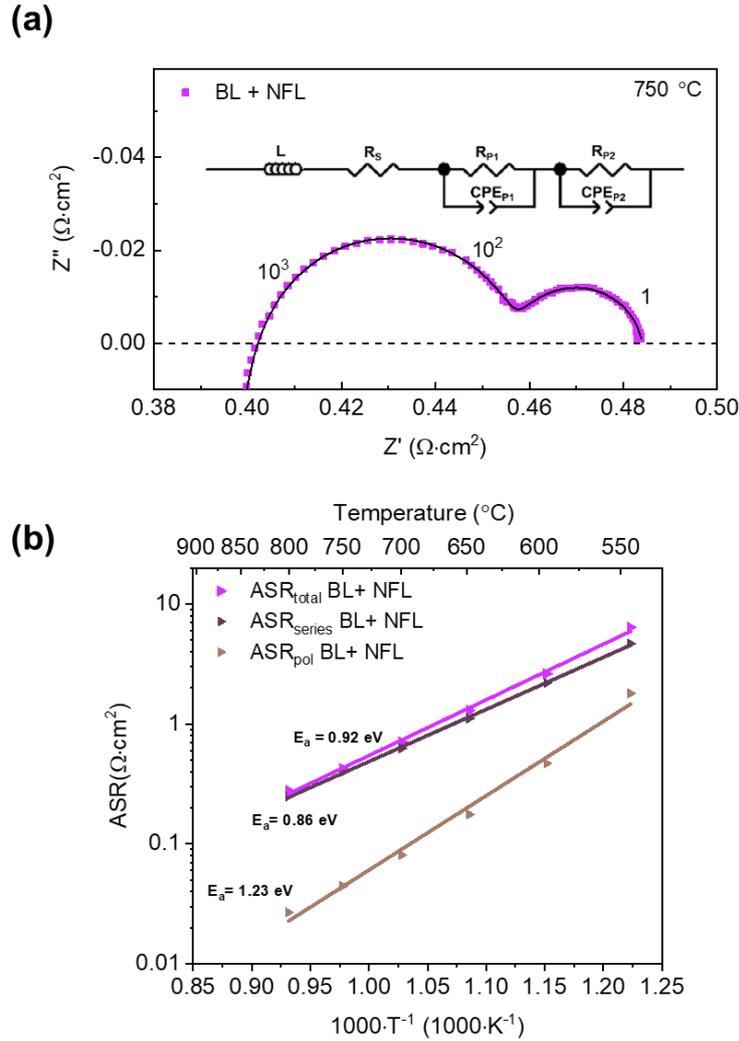

**Fig. 3 (a)** Nyquist plot of the symmetrical cells with BL + NFL and LSCF sintered at 1000 °C measured at 750 °C. **(b)** Arrhenius plot of the total, series, and polarization area specific resistances of the symmetrical cell with BL+NFL and LSCF sintered at 1000 °C.

To evaluate the electrochemical performance of the fuel cell, an anode-supported single cell with the bilayer (BL + NFL) and LSCF sintered at 1000 °C was tested. **Fig. 4 (a)** shows the current density-voltage (I-V) and current density-power density (I-P) characteristics of the anode-supported cell at 750 °C under a flow of synthetic air and dry hydrogen. An OCV of 1.14 V was measured, corresponding to the value expected for a good gas tightness between the anodic and cathodic chambers. The maximum power density of the cell was 1.07 W·cm$^{-2}$ at 0.6 V, representing a drastic performance enhancement compared with a state-of-the-art cell with configuration Ni-YSZ/YSZ/CGO/LSCF, which exhibited a power density of 0.64 W·cm$^{-2}$ at 0.6 V.[41] Also, this result is in the range of values obtained with a CGO barrier layer deposited by PLD.[20] However, it is worth mentioning that Morales et al.[20] obtained such high current densities for an annealed PLD barrier layer whereas no additional sintering step (other than the cathode sintering) was applied for the BL+NFL here presented. From the EIS diagrams at **Fig. 4 (b)** it is noted that there is a strong reduction in the polarization resistance from the sample at OCV compared to the EIS sample under polarization at current density of 0.5 A·cm$^{-2}$. This difference is mainly because the impedances were obtained in pure $H_2$ atmosphere, resulting in a complete absence of $H_2O$ at OCV, while steam internally produced under polarization reduces the resistance contributions. **Table 1** exhibits the fitting results, serial and polarization resistances and capacitances, obtained from the EIS plots of **Fig. 4 (b)**. It is noted that under bias the resistances of both polarization arcs at high and low frequency ($R_{P1}CPE_{P1}$ and $R_{P2}CPE_{P2}$) decrease while $R_S$ is independent of the applied current density. A remarkably low serial resistance is obtained (∼ 0.04 Ω·cm$^2$) thanks to the enhancement of the cathode/electrolyte interface provided by the dense BL and NFL.[20,42]

At OCV, the high frequency arc formed at ∼ 10$^3$ Hz has a contribution of ∼ 30% of the total resistance of the cell. Considering the obtained values of capacitance $C_{P1}$ and the characteristic frequencies, the resistance at the high frequency arcs can be ascribed to the charge transfer process at the electrode/electrolyte interfaces and to the oxygen reduction reaction.[37,43] Under certain polarization, these mechanisms are enhanced, as the high frequency contribution decreases by a factor of 5, from 0.53 Ω·cm$^2$ at OCV to 0.10 Ω·cm$^2$ under polarization. The EIS diagram shows that the arc formed at low frequency of ∼ 1 Hz and with $C_{P2}$ ∼ 10$^{-1}$ F·cm$^{-2}$ and $R_{P2}$ = 1.12 Ω·cm$^2$, at OCV, has the greatest contribution of ∼ 66% to the total resistance of the cell. This characteristic frequency and capacitance is typically attributed to mass diffusion mechanisms.[37,43,44] In anode supported cells that phenomena are typically dominated by gas diffusion and especially by the limited diffusion of $H_2O$. Under bias this contribution has a 10-fold decrease.

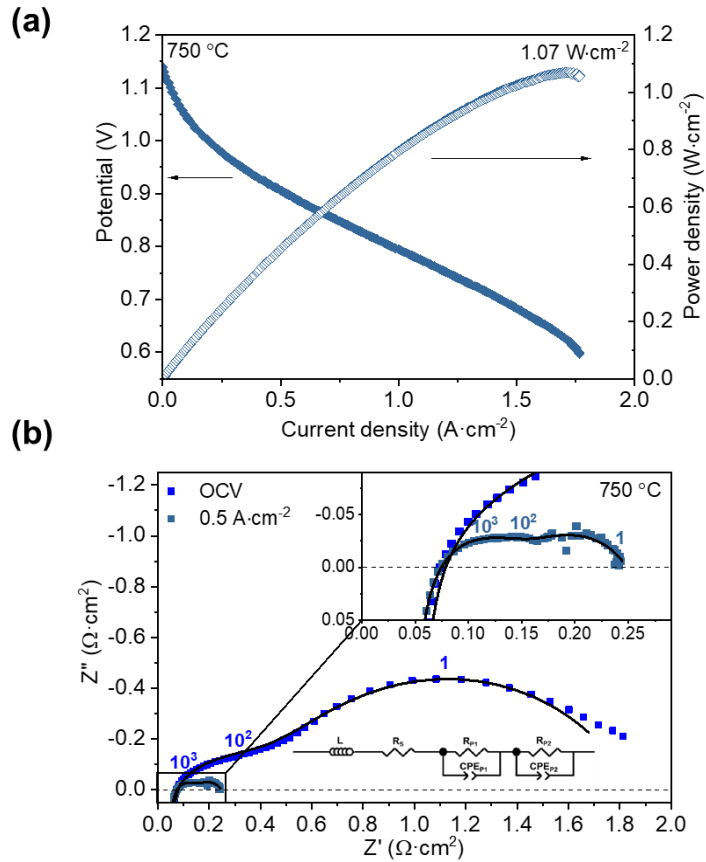

**Fig. 4 (a)** I-V and I-P curve and **(b)** complex impedance diagrams at OCV and at 0.5 A·cm$^{-2}$ of the anode-supported single cell with the BL+ NFL bilayers and LSCF at 750 °C under dry hydrogen with flow rate of 22.22 Nml·min$^{-1}$·cm$^{-2}$ and synthetic air.

Finally, to investigate the effect of the interlayers on the long-term electrochemical performance, a durability test was performed at 750 °C under dry hydrogen and synthetic air. Current densities and operation potentials during the durability test of the cell for 1500 h are shown in **Fig. 5**. Firstly, the cell operated during 250 h in the high-current output condition of 1 A·cm$^{-2}$ equivalent to 0.75 V, during which it exhibited a degradation of 190.1 mV·kh$^{-1}$ (25% V$_{in}$·kh$^{-1}$). Afterwards, milder conditions (0.5 A·cm$^{-2}$, equivalent to a potential of 0.77 V) were applied for the continuation of the durability test. At 0.5 A·cm$^{-2}$ the fuel cell operated for 1250 h with remarkable stability, with a small degradation rate of 21.6 mV·kh$^{-1}$ (2.8% V$_{in}$·kh$^{-1}$). During the durability tests I-V curves and EIS data were collected to monitor the electrochemical properties of the fuel cells, as shown in **Fig. 6**.

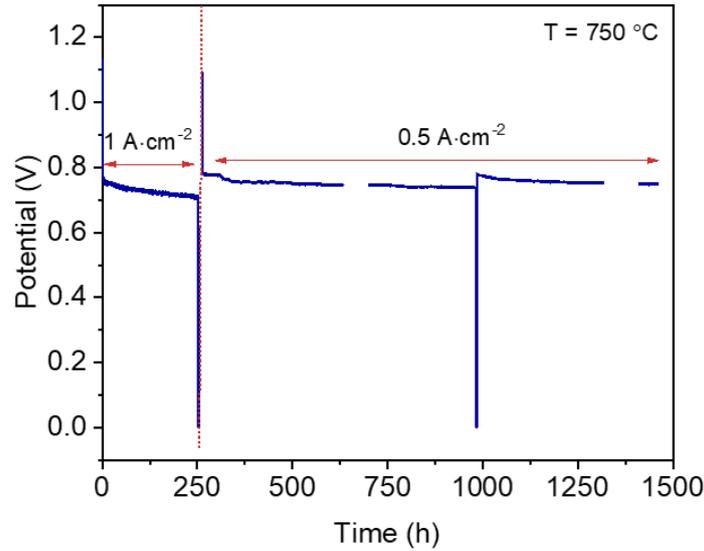

**Fig. 5** Long-term durability test of the anode-supported single cell with the BL+NFL interlayers and LSCF for ca. 1500 h at 750 °C, a current density of 1 A·cm$^{-2}$ in the initial 250 h and 0.5 A·cm$^{-2}$ for the remaining hours, under dry hydrogen with a flow density of 22.22 Nml·min$^{-1}$·cm$^{-2}$ and synthetic air.

**Fig. 6 (a)** Compares the I-V and I-P curves measured during the stability test of the anode-supported cell at 750 °C. A slight decrease of the OCV from 1.14 V to 1.09 V was observed in the initial 250 h of the test. Such OCV value remained stable until the end of the durability test. The OCV decrease indicates the presence of small potential leakages which might have appeared under more aggressive conditions applied during the first 250 h. In this first region, the power density evolved from 1.00 W·cm$^{-2}$ at 0.7 V to 0.52 W·cm$^{-2}$. When the current output was decreased to 0.5 A·cm$^{-2}$ in the second part of the durability test (milder condition), the maximum power density observed on the I-P curve evolves to 0.45 W·cm$^{-2}$. Such is compatible with the degradation observed in **Fig. 5**, which is more pronounced in the first 250 h of the test under 1 A·cm$^{-2}$ than in the following 1250 h in which the performance becomes more stable.

The EIS diagrams of the sample taken at 0 h, 250 h and 1500 h at OCV are shown in **Fig. 6 (b)** and **Table 1** exhibits the fitting results, serial and polarization resistances and capacitances, obtained from fitting the impedance plots from Fig. 6 (b). First, the serial resistance increased along with the durability test even though the final value was still low (< 0.1 Ω·cm$^2$). An increase of 37% of $R_s$ is observed after the first 250 h, followed by a total increase of 48% occurring over 1250 h. The high frequency arc, at ~ 10$^3$ Hz, shows an increase of $R_{P1}$ together with a decrease of the capacitance $C_{P1}$ with operation time. The EIS data has shown that the largest contribution to the polarization resistance originated from an increase of the resistance at a high frequency range, related with the charge transfer processes in the electrode. The arc observed at a low frequency of ~ 1 Hz ($R_{P2}CPE_{P2}$) drastically decreases after the first 250 h, but then slightly increases during the next 1250 h from 0.43 to 0.56 $\Omega \cdot cm^2$ while its capacitance ($C_{P2}$) of ~ $10^{-1}$ $F \cdot cm^{-2}$ was maintained. As previously mentioned, this arc is associated with gas diffusion, probably more related to the anodic processes, since it involves a thicker, porous layer. A decrease of this arc after the first part of the durability test is in accordance with the observed change in OCV, meaning that there was a higher pH2O the anode side. From the Nernst equation, the increase can be estimated from 1.5 v% to 4.5 v% (theoretical values of pH2O necessary to obtain an OCV of 1.14 V and 1.09 V). The logarithmic behaviour of the Nernst potential as a function of the $pH_2O$ has though, a strong impact on the polarization resistance and can hinder any microstructural degradation related to the gas diffusion. The further increase in $R_{P2}$ after the second part of the durability test can be this time attributed to a slight degradation of the cell, as the OCV did not evolve in the meantime.

To better investigate the contributions that led to the degradation of the cell, the oxidant at the cathode side was changed from synthetic air to pure $O_2$, whilst maintaining the same flow. Corresponding I-V and I-P curves are plotted in **Fig 6 (a)** and the corresponding EIS under O2 at OCV and at 0.5 $A \cdot cm^{-2}$ are plotted in Supplementary **Fig. S4**. The cell improved by 16 % under oxygen when comparing the I-V and I-P characteristics after 1500 h of operation under both atmospheres. From the impedance diagram of **Fig. S4**, it is observed that there is a decrease in the total polarization resistance, from 1.38 $\Omega \cdot cm^2$ under synthetic air to 0.65 $\Omega \cdot cm^2$ under oxygen flow at OCV. The change in gas composition affected both high and low frequency arcs, emphasizing the cathode contribution to the total resistance of the cell. Interestingly, under bias, the high frequency arc is not affected by the change of cathodic atmosphere whereas the lower frequency arc decreases when switching from air to $O_2$. This result allows understanding that both electrodes seem to have suffered degradation during the test.

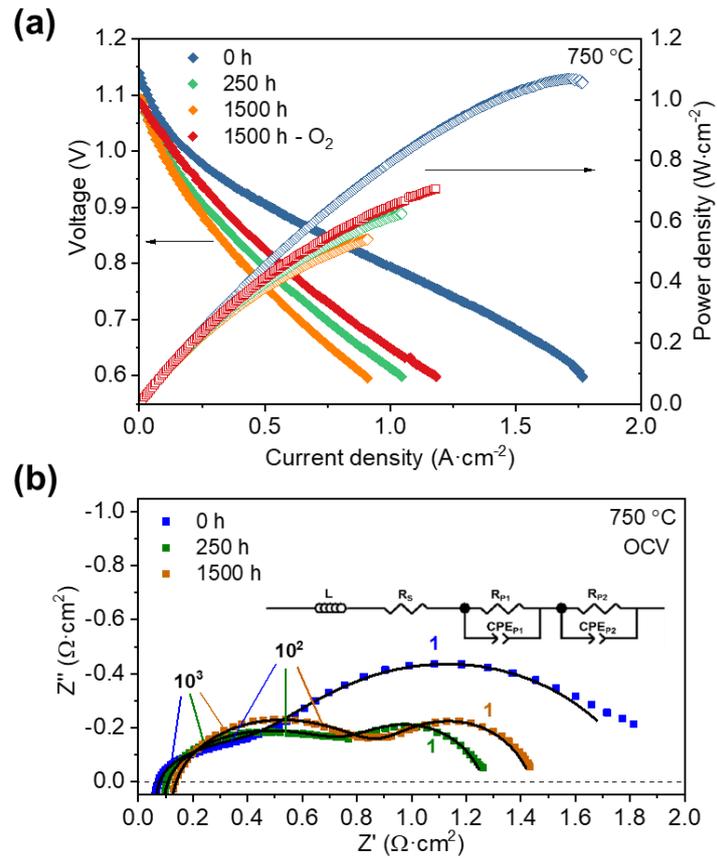

**Fig. 6 (a)** I-V and I-P curves and **(b)** complex impedance diagrams at OCV of the anode-supported single cell with the BL+NFL bilayer and LSCF at 750 °C under dry hydrogen with flow rate of 22.22 Nml·min$^{-1}$·cm$^{-2}$ and synthetic air before (0 h), after 250 h and 1500 h of operation under synthetic air on the cathode chamber (I-V and I-P curve under $O_2$ on the cathode chamber is shown in (a)).

**Table 1**. Resulting fitting parameters of series resistances ($R_S$), polarization resistances ($R_{P1}$ and $R_{P2}$) and capacitances ($C_{P1}$ and $C_{P2}$) of the anode-supported single cell with the BL+ NFL bilayer before the durability test at 0 h, after 250 h and 1500 h of operation, at OCV and at 0.5 A·cm$^{-2}$ (*plot available on **Supplementary Fig. S4**).

| Condition T = 750 °C | $R_s$ ($\Omega \cdot cm^2$) | $R_{P1}$ ($\Omega \cdot cm^2$) | $C_{P1}$ (F·cm$^{-2}$) | $R_{P2}$ ($\Omega \cdot cm^2$) | $C_{P2}$ (F·cm$^{-2}$) |
|---|---|---|---|---|---|
| 0h-OCV | 4.14E-02 | 5.29E-01 | 3.07E-03 | 1.12E+00 | 1.66E-01 |
| 0h-0.5A·cm$^{-2}$ | 4.32E-02 | 9.98E-02 | 1.04E-03 | 1.17E-01 | 1.19E-01 |
| 250h-OCV | 5.69E-02 | 7.84E-01 | 1.19E-03 | 4.32E-01 | 2.30E-01 |
| 1500 h-OCV | 8.43E-02 | 8.15E-01 | 6.07E-04 | 5.60E-01 | 1.67E-01 |
| 1500 h-OCV-O$_2$* | 8.67E-02 | 4.99E-01 | 4.30E-04 | 1.50E-01 | 2.86E-01 |
| 1500 h-0.5A·cm$^{-2}$ * | 8.25E-02 | 2.73E-01 | 3.22E-04 | 5.01E-02 | 4.78E-01 |
| 1500 h-0.5A·cm$^{-2}$-O$_2$* | 8.04E-02 | 2.09E-01 | 2.80E-04 | 4.54E-02 | 2.82E-01 |

Microstructural analysis of the fuel cell was conducted to further understand the degradation phenomena occurring particularly on the studied bilayer. SEM images of the operated cell are presented in **Fig 7**. It can be seen in **Fig 7 (a)** that the electrolyte interfaces with the anode and the cathode are well preserved with no visible delamination. A closer view (**Fig 7 (b)**) shows good adhesion between the electrolyte, functional bilayers, and cathode. However, some specific regions were found to be damaged, as seen in **Fig 7 (c)** and **(d)**. In those areas, the electrolyte presents porosity on most of the grain boundaries and grain surfaces, all along the layer thickness. The bilayer and the cathode show the formation of pores at the interface BL/electrolyte as well as inside the NFL and the cathode. Segregated nanoparticles also appeared on top of the cathode layer, corresponding most likely to promoted decomposition of the LSCF phase. The presence of pores and nanoparticles was confirmed across the whole cathode. The fast voltage decrease under harsh operating conditions (1 A·cm$^{-2}$) and the sudden voltage drop at 250 h are coherent with such an evident structural evolution leading to the occurrence to local high overpotentials and decomposition of the different oxides composing the cell.[45] These aggressive conditions could then trigger the decomposition of the cathode material and to the formation of secondary phase precipitates. The non−homogeneous distribution of the damaged regions likely come from a non−homogeneous current distribution. Degradation of the anode was not clearly observed on the microstructural characterization but phenomena such as Ni agglomeration resulting in a loss of TPB close to the electrolyte cannot be excluded.[46,47] At the light of those observations, one can conclude that the nanometric bilayer does not seem to be the main source of the observed cell degradation, but the inhomogeneity of the high current density introduced when operated at 1 A·cm$^{-2}$.

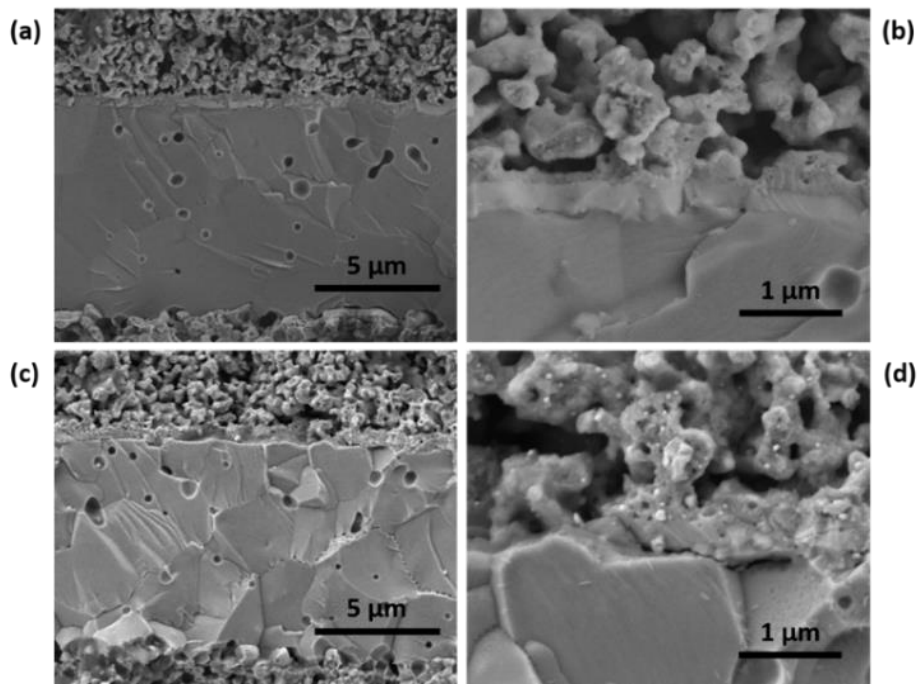

**Fig. 7** SEM micrographies of the cell operated during 1500 h **(a)** general view of the extended microstructural state of the cell and **(b)** detail of the NFL interface. **(c)** General view of the most affected zones punctually observed on the cross-section **(d)** detail of the NFL affected zone.

To complete the post-test evaluation of the BL+NFL, TEM analysis with EDX mapping around the BL+NFL interlayers was carried out on pristine and tested cells (700 °C, 0.5 A·cm$^{-2}$). In the TEM image and on the corresponding EDX analysis of the cell as prepared, **Fig. 8 (a)**, the good adhesion and uniformity of the layers and a clear distinction between the elements of each layer is clearly observed. After 700 h of operation, the microstructure of the layers is maintained. Most importantly, EDX highlights no cationic interdiffusion between cathode and electrolyte (cf. La, Zr and Sr signals), confirming that the BL+NFL bilayer is an efficient cation diffusion barrier. Note in **Fig. 8 (b)** that, in correspondence of isolated spots where the functional layer is interrupted, Sr diffusion in the electrolyte is evident, further corroborating the important role of the nanofunctional layer. Such isolated spots may derive from mechanical stress or from the presence of surface impurities during fabrication and are not expected to have an impact on the functionality given their low density (cf. **Fig. 7**). Although some cation diffusion could in principle be inferred, the authors consider that there are no significant differences on the Sr signal/noise ratios when compared with the EDX mappings of the pristine sample (**Fig. 8 (a)** and **Fig. 8 (b)**). In the literature, there is a well-described mechanism of activation of the LSM cathode by cathodic polarization which occurs by surface modification that leads to an increase of the oxygen vacancies and facilitates ORR.[48] The activation is partially due to the reduction $Mn^{3+/4+}$ to $Mn^{2+}$, which cations spread to the electrolyte surface under biased providing high electronic conductivity in the zirconia surface.[49] Interestingly, spontaneous cationic intermixing in LSM-SDC nanocomposites, determining the formation of a medium-entropy compound, was directly put in relation with improved structural stability (suppressed Sr segregation in LSM), according to recent investigations.[32] However, it is noted that cations rearrangement under bias and through the operation of the cell can have a positive effect on the durability of the cell, not only be the cause of degradation.[36]

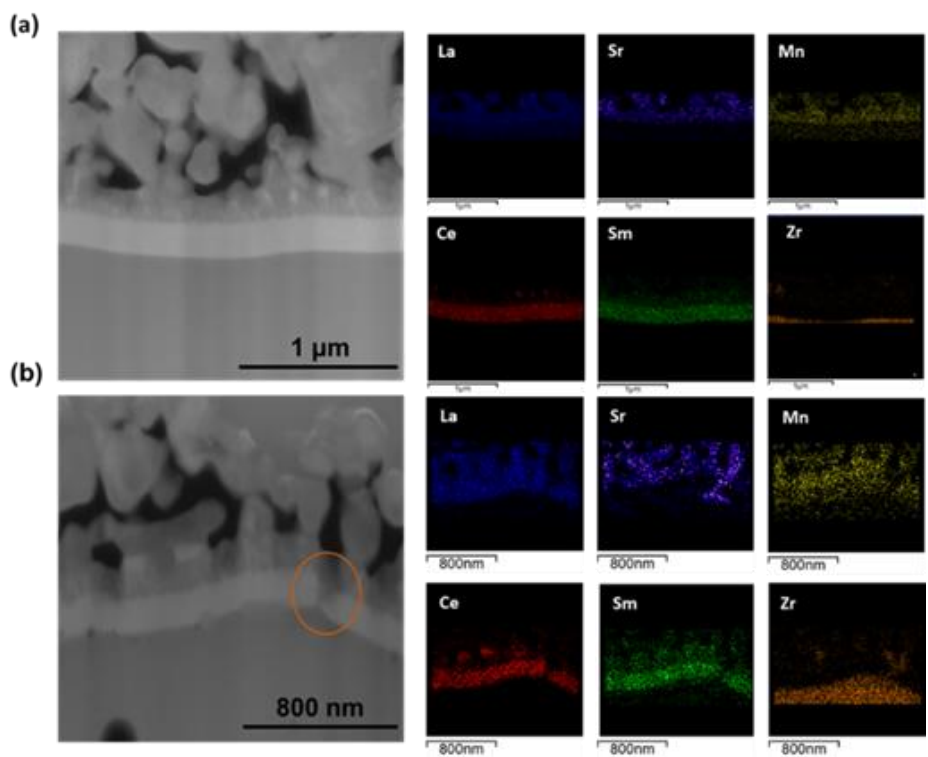

**Fig. 8** TEM images and EDX mapping of the anode-supported single cell with the BL+NFL and LSCF cathode (a) before and (b) after 700 h of operation at 0.5 A.cm$^{-2}$ at 750 °C.

## Conclusions

Bilayers of Sm-doped ceria and LSM-SDC self-assembled nanocomposites were successfully fabricated by PLD as functional layers for the electrolyte/cathode interface. The thin layers (200 nm thick) showed a dense and homogenous structure. The LSM-SDC nanocomposite layer exhibited an enhanced length of the triple phase boundary by the vertical periodicity of the phases. The combination of a barrier layer with a functional cathode layer resulted in enhanced properties of the fuel cell. An efficient barrier hindered the formation of $SrZrO_3$ during the cell fabrication with no additional annealing step required before the LSCF cathode deposition, resulting in a remarkably low polarization resistance (0.08 $\Omega \cdot cm^2$ at 750 °C), as inferred in symmetrical cells. SOFC cells using the bilayers exhibited a high-power density of 1.00 $W \cdot cm^{-2}$ at 0.7 V and 750 °C, corresponding to a significant improvement over state-of-the-art cells (~ 0.65 $W \cdot cm^{-2}$ in the same conditions) without the nanostructured bilayer. The durability of the cell was studied for 1500 h. The highest degradation occurred in the first 250 h (25% $V \cdot kh^{-1}$) when 1 $A \cdot cm^{-2}$ was drained. On the other hand, a remarkably stable performance (degradation rate 2.8% $V \cdot kh^{-1}$) was measured during 1250 h under 0.5 $A \cdot cm^{-2}$. Electrochemical impedance measurements during the stability tests indicate that the main contribution to the increased polarization resistance is associated to high-frequency arc, usually attributed to the charge transfer at electrodes, whereas the low frequency arc, ascribed to mass transport, is practically unchanged. Therefore, the degradation can be attributed to a degradation of the electric charge transport in the cell. Investigations of microstructural evolution show that the cell seemed mostly preserved with the thin nanostructured bilayer acting as an efficient diffusion barrier. However, Sr diffusion was detected in small portions of the interface, in which the defective functional bilayers showed some cracks. Such cracks may be caused during the durability tests under high current output and probably correspond to high overpotentials regions resulting in the destabilization of the engineered cathode/electrolyte interface.

The combined results show that the combination of a thin and dense barrier layer with a nanostructured cathode functional layer is an efficient strategy to engineer interfacial layers for high-performance and durable solid oxide fuel cells

**Acknowledgements**

Authors are grateful for the support of Fundação de Amparo à Pesquisa do Estado de São Paulo (FAPESP) grants no 2019/21159-4, 2019/04499-6, 2017/11937-4, 2014/50279-4, and 2014/09087-4. FCF is a fellow of the Brazilian CNPq. This project has received funding from the European Union's Horizon 2020 research and innovation program under grant agreement No 101017709 (EPISTORE).

**Notes and references**

# SUPPLEMENTARY INFORMATION

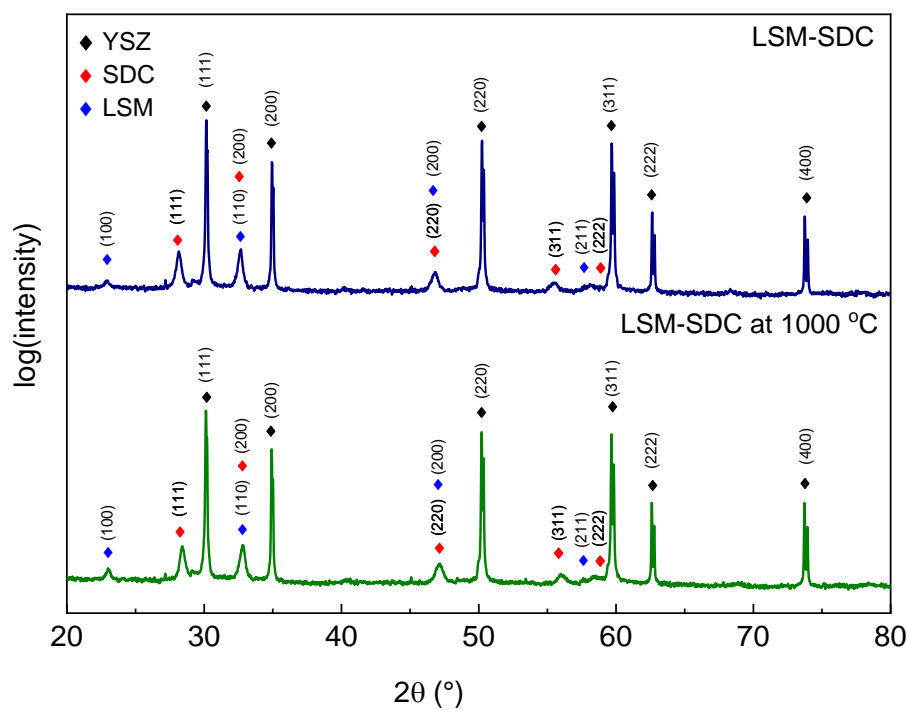

**Supplementary Fig. S1** X-ray diffraction patterns of PLD layers on YSZ. Top to bottom: LSM-SDC as-deposited, LSM-SDC after heat treatment at 1000 °C.

The electrochemical properties of the symmetrical cells with thin-film interlayers were measured by EIS to define the optimal attachment temperature of the LSCF layer through the minimisation of the ASR contributions. The EIS data of symmetrical cells with the nanocomposite LSM-SDC was collected under synthetic air in the range of 550 – 800 °C for LSCF sintering temperature of 900, 1000 and 1100 °C. **Supplementary Fig. S2 (a)** shows the impedance data of the samples at 750 °C. The experimental data was fitted with the equivalent circuit $LR_S(R_{P1}CPE_{P1})(R_{P2}CPE_{P2})$, where L refers to the inductive contribution due to the setup, $R_S$ is the serial resistance ascribed to the ionic resistance of the electrolyte and to the electronic resistance from the current collection, and the electrode polarization arcs were fitted with a resistance and a constant phase element (CPE) connected in parallel ($R_{Pi}CPE_{Pi}$). The diagrams have clearly distinguishable arcs with similar frequency distributions; the first arc at high frequency is formed at around $10^3$ and $10^2$ Hz and the low frequency arc at around 1 Hz.

The temperature dependence of the series ($ASR_{series}$) and polarization ($ASR_{pol}$) area-specific resistances for the different cells are reported in the Arrhenius plot of **Fig. S2 (b)** and **(c)** respectively. The activation energies for the $ASR_{series}$ are similar for all sintering temperatures, as seen in **Fig. 3 (a)**, but a lowest resistance is noted for the sample with the oxygen electrode sintered at temperatures of 1000 °C. From **Fig. S2 (c)**, it can be observed that the sample with attachment temperature of 1100 °C displays a significant increase in polarization resistance and a higher activation energy as compared to the samples submitted to lower temperatures. Such resistive behaviour can be attributed to the high temperature of 1100 °C that promotes the formation of resistive phases such as $SrZrO_3$ at the interface between the YSZ electrolyte and the LSCF cathode. This high temperature coupled with the limitation due to the very thin layer (200 nm) might activate parallel interdiffusion processes between the cathode and the electrolyte leading to the formation of insulating phases and decomposition of the electrode material. The sample with layers attached 1000 °C has the lowest polarization and series resistance at the cell's operating temperature (700 – 800 °C). Thus, such sintering temperature was selected as the optimal one to produce the symmetrical cell with both the NFL and the BL.

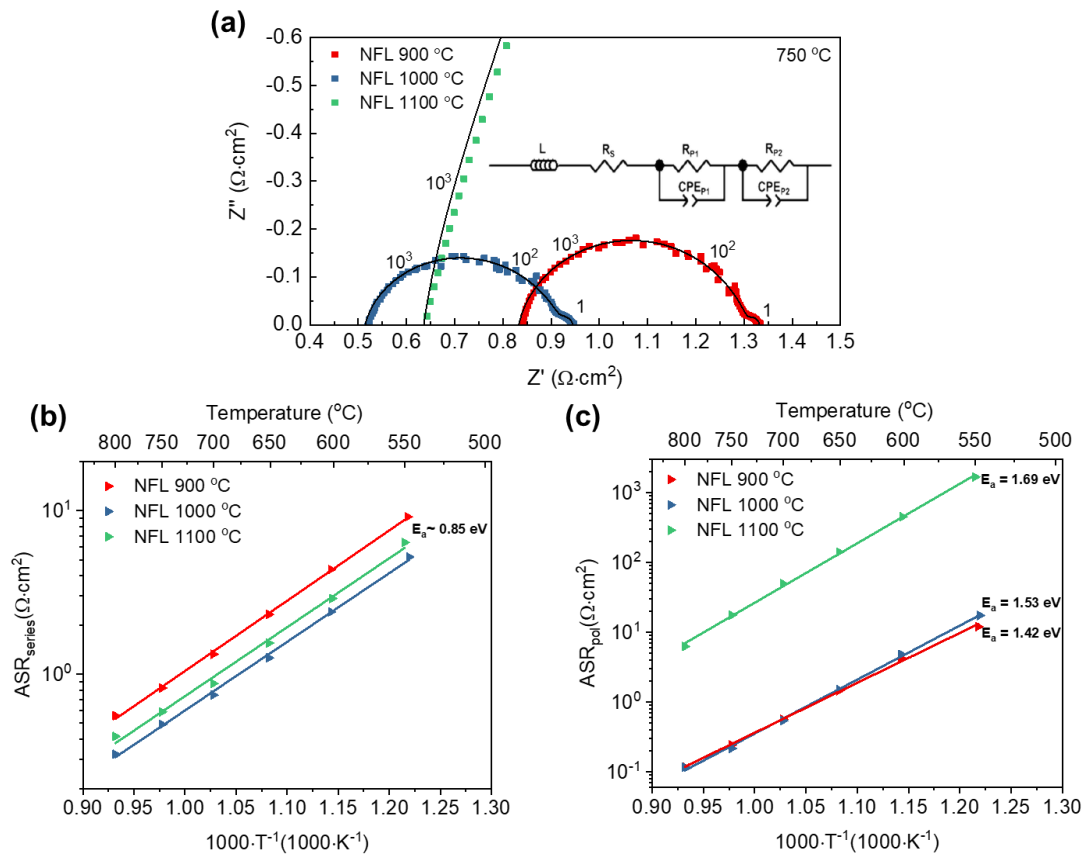

**Supplementary Fig. S2** (a) Impedance diagram of the symmetrical cells with YSZ support, NFL and LSCF cathode submitted to different sintering temperatures, 900, 1000 and 1100 °C, measured at 750 °C. Arrhenius plots of the series ASR (b) and of the polarization ASR (c) of the symmetrical cells.

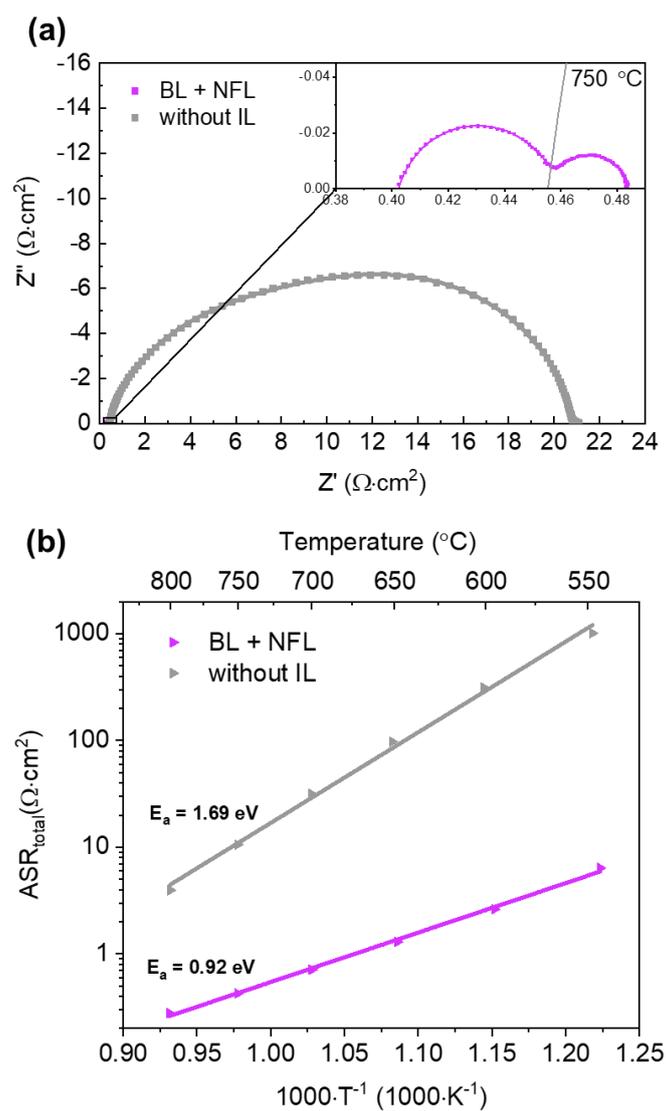

**Supplementary Fig. S3 (a)** Impedance diagram of the symmetrical cell with and without BL and NFL with LSCF sintered at 1000 °C measured at 750 °C. **(b)** Arrhenius plot of the total area specific resistance of the symmetrical cell with BL+NFL and without intermediary layers (IL) and LSCF sintered at 1000 °C.

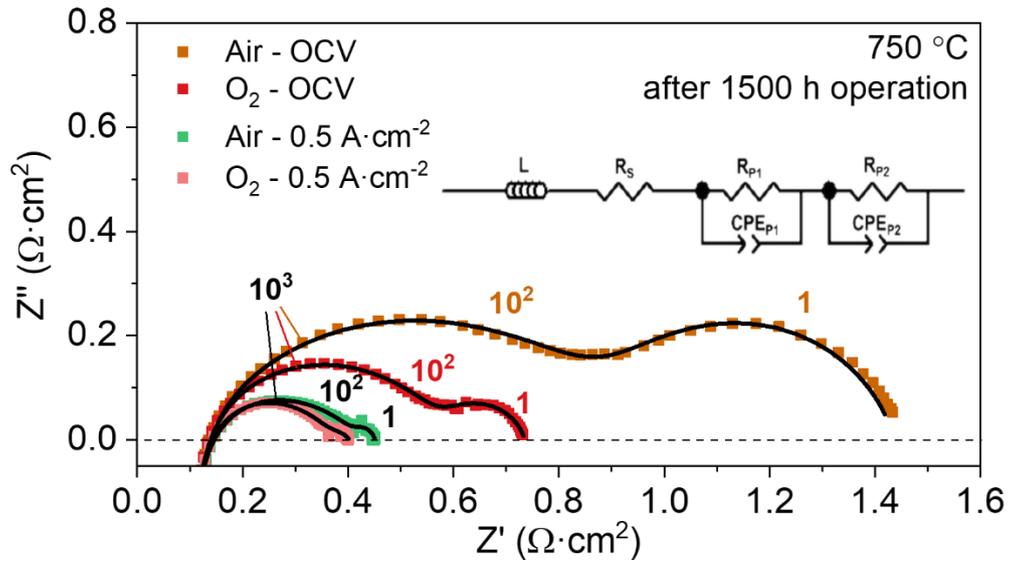

**Supplementary Fig. S4** Impedance diagram of the anode-supported single cell with the BL+NFL bilayers and LSCF at 750 °C under dry hydrogen with flow rate of 22.22 Nml·min1·cm$^{-2}$ after 1500 h of operation under synthetic air and O$_2$ on the cathode chamber at OCV and under 0.5 A·cm$^{-2}$.